\documentclass[12pt]{article}
\usepackage{palatino} 
\begin{document}
\begin{titlepage}
\begin{flushright}
\end{flushright}
\vskip.3in

\begin{center}
{\Large \bf Extended integrability r\'egime for the \\ supersymmetric U model}
\vskip.3in
{\large Jon Links} 
\vskip.2in
{\em Department of Mathematics, \\ The University of Queensland, 
      4072, \\ Australia

Email: jrl@maths.uq.edu.au}
\end{center}

\vskip 2cm
\begin{center}
{\bf Abstract}
An extension of the supersymmetric $U$ model for correlated electrons is
given and integrability is established by demonstrating that the model
can be constructed through the Quantum Inverse Scattering Method using
an $R$-matrix without the difference property. 
Some general symmetry properties of the model are discussed and 
from the Bethe ansatz solution an expression for the energies is 
presented.  
\end{center}

\vskip 3cm

\end{titlepage}


\def\a{\alpha}
\def\b{\beta}
\def\d{\delta}
\def\e{\epsilon}
\def\ve{\varepsilon}
\def\g{\gamma}
\def\k{\kappa}
\def\l{\lambda}
\def\n{\nu} 
\def\o{\omega}
\def\t{\theta}
\def\si{\sigma}
\def\D{\Delta}
\def\L{\Lambda}
\def\X{\bar{X}}
\def\Y{\bar{Y}}
\def\Z{\bar{Z}} 
\def\ch{\check}
\def\f{{\cal F}} 
\def\G{{\cal G}}
\def\hG{{\hat{\cal G}}}
\def\R{{\cal R}}
\def\hR{{\hat{\cal R}}}
\def\C{{\bf C}}
\def\P{{\bf P}}
\def\T{{\cal T}}
\def\H{{\cal H}}
\def\trho{{\tilde{\rho}}}
\def\tphi{{\tilde{\phi}}}
\def\tT{{\tilde{\cal T}}}
\def\uqsnh{{U_q[\widehat{sl(n|n)}]}}
\def\uqs1h{{U_q[\widehat{sl(1|1)}]}}
\def\ot{\otimes}


\def\beq{\begin{equation}}
\def\eeq{\end{equation}}
\def\bea{\begin{eqnarray}}
\def\eea{\end{eqnarray}}
\def\ba{\begin{array}}
\def\ea{\end{array}}
\def\no{\nonumber}
\def\lt{\left}
\def\rt{\right}
\def\dag{\dagger}
\def\upa{\uparrow}
\def\doa{\downarrow} \newcommand{\bq}{\begin{quote}}
\newcommand{\eq}{\end{quote}}

\newtheorem{Theorem}{Theorem}
\newtheorem{Definition}{Definition}
\newtheorem{Proposition}{Proposition}
\newtheorem{Lemma}{Lemma}
\newtheorem{Corollary}{Corollary}
\newcommand{\proof}[1]{{\bf Proof. }
        #1\begin{flushright}$\Box$\end{flushright}}

The supersymmetric (SUSY) $U$ model was first introduced in \cite{bglz} as an
example of a system of correlated electrons which is integrable in one
dimension as a consequence of the Quantum Inverse Scattering Method 
(QISM) (e.g. see \cite{fst}). 
Such models, which can be solved exactly by the Bethe
ansatz method, are important in that the exact solutions 
offer non-perturbative results concerning physical behaviour. 
For the SUSY $U$ model, Bethe ansatz
solutions have been studied \cite {m,bf,bkz,hgl,rm,pf1,pf2} and 
several analyses into
the physical characteristics that the model describes have been undertaken
\cite{bf,bkz,rm,jks}. 

The construction of the SUSY $U$ model is based on an $R$-matrix
satisfying the Yang-Baxter equation associated with the one-parameter
family of minimal typical representations of the Lie superalgebra
$gl(2|1)$. 
In terms of the standard notation for electron creation, annihilation
and occupation operators the local (two-site)
Hamiltonian for the model reads
\bea
h_{i,i+1}&=&-\sum_{\si}(c^{\dag}_{i\si}c_{i+1\si} +h.c.)
(1+U)^{1/2(n_{i,-\si}+n_{i+1,-\si})} \no     \\
&&     +U(c_{i_\doa}^{\dag}c_{i\upa}^{\dag}c_{i+1\upa}
c_{i+1\doa} +h.c.)   \no  \\
&& +U(n_{i\upa}n_{i\doa} +n_{i+1\upa}n_{i+1\doa})-2  \no \\
&& +n_{i\upa}+n_{i\doa}+n_{i+1\upa}+n_{i+1\doa},
\no \eea
where $U$ is an arbitrary free parameter. The local 
Hamiltonian is also invariant with
respect to the Lie superalgebra $gl(2|1)$ (hence the name). Below, an
extension of this model will be derived in such a way that integrability
is  maintained. The local Hamiltonian of the new model reads
\bea
h_{i,i+1}&=&-\sum_{\si}\lt((1-it)c^{\dag}_{i\si}c_{i+1\si} +h.c.\rt)
(1+U)^{1/2(n_{i,-\si}+n_{i+1,-\si})} \no   \\
&&     +U\lt((1-it)c_{i_\doa}^{\dag}c_{i\upa}^{\dag}c_{i+1\upa}
c_{i+1\doa} +h.c.\rt)   \no  \\
&& +U(n_{i\upa}n_{i\doa} +n_{i+1\upa}n_{i+1\doa})-2 \no \\
&& +n_{i\upa}+n_{i\doa}+n_{i+1\upa}+n_{i+1\doa} 
\label{tu} \eea
where now $t$ is an additional free variable which when chosen to be
real (along with $U$ real and $U>-1$) results in a Hermitian Hamiltonian.
For the case $t=0$ the usual SUSY $U$ model is recovered. 
The extended 
model bears some similarity with the multiparametric SUSY $U$ model
constructed in \cite{flr} but is in fact inherently different.

The construction of the above model is through the use of a solution
of the Yang-Baxter equation without difference property in the spectral
parameter. It is known that the  Hubbard model may be derived via
the QISM using an $R$-matrix which is also  without the difference
property \cite{sh,mr}. 
However, for the Hubbard model the Lax operator is given as a particular
coupling of two auxilliary Lax operators of six-vertex type. 
The construction employed here appears 
more akin to the generalized chiral Potts models given in \cite{djmm}
based on representations of quantum algebras at roots of unity. For
these models and the one discussed here the spectral parameters without
difference property originate from  the {\it representation} of the underlying
algebraic structure.

In  order to demonstate integrability of this model, we begin with
the rational limit of the $U_q(gl(2|1))$ invariant (i.e. $gl(2|1)$
invariant) solution of the Yang-Baxter equation
with additional spectral paramters constructed in \cite{bdgz,dglz1}. This 
solution may be written in the form 
\beq R(u,\b,\a)=\frac{u-\a-\b}{u+\a+\b}P_1+P_2+\frac{ u+\a+\b+2}{  
u-\a-\b- 2}P_3 \label{rm} \eeq 
and satisfies the Yang-Baxter equation 
$$ R_{12}(u-v,\b,\g)R_{13}(u,\b,\a)R_{23}(v,\g,\a)
=R_{23}(v,\g,\a)R_{13}(u,\b,\a)R_{12}(u-v,\b,\g).$$ 
Note that this solution of the Yang-Baxter equation has multiple
spectral parameters, one of which has the difference property while the
other two do not. It is worth remarking at this point that the above
solution is just one of a plethora of solutions of this type which arise
naturally as a consequence of the representation theory of the type I 
quantum superalgebras \cite{dglz1,dglz2}. 

In the above expression for the $R$-matrix the operators $P_i$ are
$gl(2|1)$ invariant projection operators. To explain their actions we
begin by recalling that $gl(2|1)$ has 
 generators $E^i_j,\,i,j=1,2,3$ satisfying the
super commutator relations
$$[E^i_j,\,E^k_l]=\delta^k_jE^i_l-(-1)^{([i]+[j])([k]+[l])}\delta^i_l
E^k_j. $$
Above, $[1]=[2]=0,\,[3]=1.$ Choosing a four dimensional space with basis
$\{\left|i\right>:\,i=1,2,3,4\}$, a representation $\pi_{\a}$ 
of $gl(2|1)$ acting
on this space exists with the action of the generators given by
\vfil\eject 
\begin{eqnarray}
&&E^1_2=\left|2\right>\left<3\right|,\nonumber \\
&&E^2_1=\left|3\right>\left< 2\right|,\nonumber \\
&&E^1_1=-\left|3\right>\left< 3\right|-\left|4\right>\left<
4\right|,\nonumber \\
&&E^2_2=-\left|2\right>\left< 2\right|-\left|4\right>\left<
4\right|,\nonumber \\
&&E^2_3=\sqrt{\alpha}\,\left|1\right>\left<
2\right|+\sqrt{\alpha+1}\,\left|3\right>\left< 4\right|,\nonumber \\
&&E^3_2=\sqrt{\alpha}\,\left|2\right>\left<
1\right|+\sqrt{\alpha+1}\,\left|4\right>\left< 3\right|,\nonumber \\
&&E^1_3=-\sqrt{\alpha}\,\left|1\right>\left<
3\right|+\sqrt{\alpha+1}\,\left|2\right>\left< 4\right|,\nonumber \\
&&E^3_1=-\sqrt{\alpha}\,\left|3\right>\left<
1\right|+\sqrt{\alpha+1}\,\left|4\right>\left< 2\right|,\nonumber \\
&&E^3_3=\alpha\,\left|1\right>\left< 1\right|+(\alpha+1)\,\left
(\left|2\right>\left< 2\right|+\left|3\right>\left< 3\right|\right )
  +(\alpha+2)\,\left|4\right>\left<4\right|.\no 
      \end{eqnarray}
Above, the states $\left|1\right>,\,\left|4\right>$ are bosonic and
$\left|2\right>,\,\left|3\right>$ are fermionic. The highest weight
state is $\left|1\right>$ with weight $(0,0|\alpha)$. It is
significant here that $\alpha$ is a free complex parameter. Considering
the tensor product representation $\pi_{\b}\ot \pi_{\a}$ 
for generic values of $\a$ and $\b$, then $P_1$
projects onto the irreducible submodule with (unnormalized) basis
vectors 
\begin{eqnarray}
&&\left|\Psi^1_1\right>=\left|1\right>\otimes \left|1\right>,\nonumber
\\
&&\left|\Psi^1_2\right>=\sqrt{\b}\left|2\right>\otimes
\left|1\right>+
\sqrt{\a}\left|1\right>\otimes \left|2\right>,\nonumber \\
&&\left|\Psi^1_3\right>=\sqrt{\b}\left|3\right>\otimes
\left|1\right>+
\sqrt{\a}\left|1\right>\otimes \left|3\right>,\nonumber \\
&&\left|\Psi^1_4\right>=\sqrt{\b(\b+1)}\left|4
\right>\otimes \left|1\right>
+\sqrt{\a(\a+1)}\left|1\right>\otimes
\left|4\right>)+\sqrt{\alpha\b}(\left|2\right>\otimes
\left|3\right>-\left|3\right>\otimes \left|2\right>),\nonumber \\
\eea 
while $P_3$ projects onto the irreducible space spanned by 

\bea 
&&\left|\Psi^2_1\right>=\sqrt{\a(\a+1)}\left|4
\right>\otimes \left|1\right>
+\sqrt{\b(\b+1)}\left|1\right>\otimes
\left|4\right>) \no \\
&&~~~~~~~~~~~~~+\sqrt{(\a+1)(\b+1)}(-\left|2\right>\otimes
\left|3\right>
+\left|3\right>\otimes
\left|2\right>),\nonumber \\
&&\left|\Psi^2_2\right>=\sqrt{\b+1}\left|2\right>\otimes
 \left|4\right>+
\sqrt{\a+1} \left|4\right>\otimes
 \left|2\right>),\nonumber \\
&&\left|\Psi^2_3\right>=\sqrt{\b+1}\left|3\right>\otimes
 \left|4\right>+
\sqrt{\a+1}\left|4\right>\otimes
\left|3\right>),\nonumber\\
&&\left|\Psi^2_4\right>=\left|4\right>\otimes
 \left|4\right>.
\no 
\end{eqnarray}
The projector $P_2$ is obtained from $P_2=I-P_1-P_3$. From this solution
of the Yang-Baxter equation we may now construct the transfer matrix 
$$t(u,\b,\a)={\rm tr}R_{0L}(u,\b,\a)....R_{02}(u,\b,\a)R_{01}(u,\b,\a)$$
which forms a commuting family in two variables; viz.
$$[t(u,\b,\a),\,t(v,\g,\a)]=0$$
and thus $t(u,\b,\a)$ can be diagonalized independently of both $u$ and
$\b$. In fact, the diagonalization of this transfer matrix has already
been treated in \cite{pf1} with the result that the eigenvalues are given
by 
\beq \Lambda(u,\b,\a)=\lt(\frac{u-\a-\b}{u+\a+\b}\rt)^L\overline{\Lambda}
(u,\b,\a)  \eeq 
with 
\bea\overline{\Lambda}(u,\b,\a)&=&\prod_{i=1}^n\frac{u-\l_i+\b}{u-\l_i-\b} 
\no \\ 
&&+\lt(\frac{u+\a-\b}{u-\a-\b}.\frac{u+\a-\b-2}{u-\a-\b-2}\rt)^L  
\prod_{i=1}^n\frac{u-\l_i+\b}{u-\l_i-\b-2} \no \\ 
&&-\lt(\frac{u+\a-\b}{u-\a-\b}\rt)^L\lt\{\prod_{i=1}^n\frac{u-\l_i+\b
}{u-\l_i-\b}\prod_{j=1}^m\frac{u-\n_j-\b+1}{u-\n_j-\b-1}\rt. \no \\
&&~~~~~~+\lt.\prod_{i=1}^n\frac{u-\l_i+\b}{u-\l_i-\b-2} 
\prod_{j=1}^m\frac{u-\n_j-\b-3}{u-\n_j-\b-1}\rt\}  \eea 
such that the parameters $\l_i,\,\n_j$ are solutions of the Bethe ansatz
equations 
\bea 
\lt(\frac{\l_k+\a}{\l_k-\a}\rt)^L&=&\prod_{j=1}^m\frac{\l_k-\nu_j-1}{
\l_k-\n_j+1},~~~~~~k=1,....,n,  \no \\  
\prod_{k=1}^n\frac{\l_k-\n_i+1}{\l_k-\n_i-1}&=&-\prod_{j=1}^m\frac{
\n_j-\n_i+2}{\n_j-\n_i-2},~~~~~~i=1,....,m. \no  
\eea 

The $R$-matrix possesses the property 
$$R(0,\a,\a) = -P$$ 
where $P$ is the ${\bf Z}_2$-graded permutation operator. From here
on in we make the parameterization
$$\b=itu+\a$$   
and fix $t$ and $\a$. Writing the $R$-matrix now as a function of only
$u$ we have that 
$$R(u=0)=-P$$ 
and thus by the standard approach of the QISM \cite{fst} 
we can construct a closed
periodic quantum model where the Hamiltonian is the logarithmic
derivative of the transfer matrix and expressible as 
$$H=\sum_{i=1}^{L-1}h_{i(i+1)} + h_{L1}$$ 
with the two-site Hamitonian given by 
$$h=-P.\lt.\frac{d}{du}R(u)\rt|_{u=0}.$$ 
In terms of the projections operators (cf. (\ref{rm})) we
have 
$$h=\frac{-1}{\a}P_1+\frac{1}{\a+1}P_3-2it \lt.P\frac{dP_2}{d\b}\rt|_{\b=\a}  
$$
which when expressed in terms of Fermi operators leads to (\ref{tu})
with $U=\a^{-1}$ and an overall normalization factor of $-2(\a+1)$ 
included for convenience. 
The energies of the Bethe states
are obtained from the transfer matrix eigenvalues
\bea E&=& -2(\a+1)\lt.\Lambda^{-1}\frac{d\Lambda}{du}\rt|_{u=0} \no \\
&=& \frac{2L(\a+1)}{\a}+4(\a+1)
\sum_{i=1}^n\frac{\a+it\l_i}{\l^2_i-\a^2}. \no \eea

In conclusion we discuss some remarkable properties this model
possesses.  By construction, the
 transfer matrix from which it is derived is $gl(2|1)$ invariant. In
fact, one may show that the eigenstates obtained by the algebraic Bethe
ansatz method are highest weight states in complete analogy with other
$gl(2|1)$ invariant models studied in \cite{fk,lf,flt}.
However, the local Hamiltonians are {\it not} $gl(2|1)$ invariant. Only
for the global system is the supersymmetry present. Regardless,  
it is still   
possible to add arbitrary chemical potential and magentic field terms
to the local Hamiltonian (\ref{tu}) which do not violate the
integrability. 

Construction of the
model for $L=2$ yields the usual supersymmetric $U$ model, as does the
construction on an open chain using (an appropriate modification of) 
Sklyanin's approach \cite{s} 
where the boundary $K$-matrices are chosen
to be trivial. These unusual scenarios are a consequence of the property
$$\lt.P\frac{dP_2}{d\b}\rt|_{\b=\a}=-\lt.\frac{dP_2}{d\b}\rt|_{\b=\a}P$$ 
which is precisely the symmetry breaking term for the local
Hamiltonians, and the presence of which also implies that 
$$h_{i(i+1)}\neq h_{(i+1)i}.   $$     

\vskip.3in
\noindent {\bf Acknowledgements.}

I wish to thank Angela Foerster, Ruibin Zhang and David De Wit for 
discussions and assistance. 
This work has been supported by the Australian Research Council.

\newpage

\end{document}